\documentclass[twocolumn,showpacs,preprintnumbers,amsmath,amssymb,floatfix]{revtex4}
\usepackage{graphicx}% Include figure files
\usepackage{dcolumn}% Align table columns on decimal point
\usepackage{bm}% bold math

\newcommand{\beq}{\begin{equation}}
\newcommand{\eeq}{\end{equation}}
\newcommand{\bey}{\begin{eqnarray}}
\newcommand{\eey}{\end{eqnarray}}

\begin{document}

\title{ Analytical model of strange  star in the low-mass X-ray binary 4U 1820-30  }

\author{ Mehedi Kalam}
\email{kalam@iucaa.ernet.in} \affiliation{Department of
Physics, Aliah University, Sector - V , Salt Lake,  Kolkata -
700091, India}

\author{Farook Rahaman}
\email{rahaman@iucaa.ernet.in} \affiliation{Department of
Mathematics, Jadavpur University, Kolkata 700 032, West Bengal,
India}
\author{Sajahan Molla}
\email{sajahan.phy@gmail.com} \affiliation{Department of Physics,
Aliah University, Sector - V , Salt Lake,  Kolkata, India}
\author{Md.Abdul Kayum Jafry}
\email{akjafry@yahoo.com} \affiliation{Department of Physics,
Shibpur Dinobundhoo Institution (College), Howrah 711102, West
Bengal, India}
\author{Sk.Monowar Hossein}
\email{hossein@iucaa.ernet.in} \affiliation{Department of
Mathematics, Aliah University, Sector - V , Salt Lake,  Kolkata -
700091, India}

\date{\today}

\begin{abstract}
In this article, we have proposed a model for a
realistic strange star under Tolman VII metric\citep{Tolman1939}. Here the field
equations are reduced to a system of three algebraic equations for
anisotropic pressure.
 Mass, central density and surface density of strange star in the low-mass X-ray binary 4U 1820-30
  has been matched with the observational data according to our model. Strange materials
clearly satisfies the stability condition (i.e. sound velocities  $
<1 $) and TOV-equation. Here also surface red shift of the star has
been found to be within reasonable limit.
\end{abstract}

 \maketitle

\section{Introduction}
Compact objects are of great attention for a long time. Several
researchers\citep{Rahaman2012a,Kalam2012a,Hossein2012,Rahaman2012b,Kalam2012b,Kalam2013,Lobo2006,Bronnikov2006,Egeland2007,Dymnikova2002}
investigated compact stars analytically or numerically. Stars, in
general, are evolved by burning lighter elements into heavier nuclei
from the time of birth. In the end of nuclear burnning  white-dwarf,
neutron stars, quark stars, dark stars and eventually black holes
may formed due to strong gravity.  To include the effects of local
anisotropy, Bowers and Liang (1974)\citep{Bowers1974} stressed on
the importance of local anisotropic equations of state for
relativistic fluid sphere. They showed that anisotropy may have
effects on such parameter like maximum equilibrium mass and surface
redshift. In stellar system, Ruderman (1972)\citep{Ruderman1972}
argued that, in very high density range $(\sim ~10^{15}~ gm
~cm^{-3})$ nuclear matter may have anisotropic features and nuclear
interaction should be treated relativistically. Anisotropy in matter
indicates that radial pressure $(p_{r})$ is not same as the
tangential pressure $(p_{t})$. A star becomes anisotropic, if its
matter density exceeds the nuclear
density\citep{Bowers1974,Sokolov1980, Herrera1992}. This phenomenon
may occur for existence of solid core, phase transition, presence of
electromagnetic field etc. 4U 1820-30 resides in the globular
cluster NGC 6624. It is an ultra-compact binary and has an orbital
period of 11.4
 minutes\citep{Stella1987}.  During Rossi X-ray Timing Explorer(RXTE) observations, it has been observed that 4U 1820-30 exhibits a super-burst.
 Possibly this is due to  burning of a large mass of carbon \citep{Strohmayer2002}.
The 4U 1820-30 exhibits super burst, however , these strange
stars may be made of chemically equilibrated strange matter.
Scientists are searching that
 matter distribution which should be incorporated in energy momentum tensor to describe strange stars. This paper depicts how this is accomplished mathematically
  and discuss the consequences of the properties of the strange stars. There are many high masses stars are found in different types of pulsar binaries. In these
   cases the masses rely an observation of periastron advance which is believed to be due to general relativistic effects only rather than other effects due to
   rotationally and tidally induced quardrupoles. One of the useful tool for determining masses of the compact stars is X-ray eclipses. The binary eclipses are
   approximated analytically  by assuming  that the companion star is spherical with an effective Roche lobe radius.

 In 1939, Tolman\citep{Tolman1939} proposed
 static solutions for a sphere of fluid. In that article, he pointed out that due to complexity of the VII-th   solution
  (among the eight different solutions), it is not a feasible one for physical consideration
  ( there was a  misprint   in the Tolman solution VII (4.7) but that does not affect the original solution).
It seems due to complicated nature of the solution he did not able
to provide more physical properties of the solution.
  Rather we say that he did not try to explore physics of his solution VII due to complexity of the solution.
We thought this solution may explore some physics. In this work,
we have shown that
  this solution would be interesting in the sense that this solution corresponds to the interior of strange
  stars.
  We, here, want to check the feasibility of our model by taking the Tolman solution VII.\\
 Motivated by the above fact, we are specifically interested for modeling strange star in the low-mass X-ray binary 4U 1820-30.
 We compare  our measurements of mass, radius , central density, surface density and
 surface red-shifts with  the strange star  in the low-mass X-ray binary 4U 1820-30 and it is found to be   consistent
 with standard data\citep{Guver2010}.\\
The density within the strange stars are normally beyond the
nuclear matter density. The theoretical advances in last few
decades indicate that pressures
 within the stars are anisotropic. Thus one would expect anisotropy plays a major
 role for modelling these stars.

 We have considered the interior
 space time geometry of the strange star is Tolman VII type and try to
 investigate the matter distributions which produce this space time. Our
 calculations show  that the matter distribution that produces Tolman VII type
 spacetime geometry should be anisotropic. This helps us for modelling strange
 star which is anisotropic in nature  as the density within the strange stars
 are normally beyond the nuclear matter density.

  In this work, we have chosen
   the interior
 space time geometry of the strange star is Tolman VII type and try to
 investigate the matter distributions which produce this space time.
  We have assumed only Tolman VII space time for modelling
strange stars. The other solutions of Tolman are not interesting
to us as far as we have studied \citep{Tolman1939}.  In Tolman I solution,
 $e^\nu=constant$ , i.e. redshift function is constant, therefore not interesting.
 Tolman II corresponds to Schwarzschild de-Sitter solution.
 In Tolman III solution, the energy density is constant, therefore not interesting.
 In Tolman IV and V, redshift function are very much specific, therefore not interesting.
 In Tolman VI, coefficient of $dr^{2}$  has been taken as constant. Therefore we did not
 consider.
 The Tolman VIII solution,
\begin{eqnarray}
ds^{2}=e^{-\lambda}[B^{2}r^{2b}dt^2 - dr^2 - r^2e^{\lambda}
d\theta^2 +sin^2\theta d\phi^2]\nonumber
\end{eqnarray}
is conformally related to the metric whose redshift function is very specific ( polynomial function of r ). So, we discarded  it.

 We organize our paper as follows:\\
  In Sec II, we have provided the basic equations in connection to the  Tolman VII metric. In Sec. III,
   we have studied the physical behaviors of the star namely, anisotropic behavior, Matching
conditions, TOV equations, Energy conditions, Stability
and  Mass-radius relation \& Surface redshift in different sub-sections. The article concluded
with a short discussion.

\section{Interior solution}
We assume that the interior space-time of a star is described by the metric
\begin{eqnarray}
ds^2 = -B^2\sin ^{2}\ln \sqrt{\frac{\sqrt{1-\frac{r^2}{R^2}+4\frac{r^4}{A^4}}+2\frac{r^2}{A^2}-\frac{1}{4}\frac{A^2}{R^2}}{C}}~dt^2 \nonumber \\
 +\left(1-\frac{r^2}{R^2}+4\frac{r^4}{A^4}\right)^{-1}dr^2
+r^2d\Omega^{2} ~~~~~\label{eq1}
\end{eqnarray}
where $R$, $C$, $A$, $B$ are constants. Such type of metric (1) was
proposed by Tolman \citep{Tolman1939}(known as Tolman VII metric)to develop a viable
model for a  star. We assume that the energy-momentum tensor
for the interior of the star has the standard form
\begin{equation}
               T_\nu^\mu=  (-\rho , p_{r}, p_{t}, p_{t}),
         \label{Eq2}
          \end{equation}

where $\rho$ is the energy-density, $p_r$ and $p_t$ are the radial
and transverse pressure respectively. Einstein's field equations
accordingly are obtained as ($ c=1,G=1$)
\begin{eqnarray}
8\pi  \rho &=& \left(1-\frac{r^2}{R^2}+4\frac{r^4}{A^4}\right)\left[\frac{\lambda^\prime}{r}-\frac{1}{r^2}\right]+\frac{1}{r^2},\label{eq2}\\
8\pi  p_{r} &=& \left(1-\frac{r^2}{R^2}+4\frac{r^4}{A^4}\right)\left[\frac{\nu^\prime}{r}+\frac{1}{r^2}\right]-\frac{1}{r^2},\label{eq3}\\
  ~~~ 8\pi
p_{t}&=&\frac{1}{2}\left(1-\frac{r^2}{R^2}+4\frac{r^4}{A^4}\right)\nonumber
\\&&\left[\nu^{\prime\prime}+\frac{\nu^\prime-\lambda^\prime}{r}+\frac{{{\nu^\prime}^2}-{\lambda^\prime\nu^\prime}}{2}\right]. \label{eq4}
\end{eqnarray}
where
\[ e^\lambda = \left(1-\frac{r^2}{R^2}+4\frac{r^4}{A^4}\right),\]
\[ e^\nu =B^2\sin ^{2}\ln \sqrt{\frac{\sqrt{1-\frac{r^2}{R^2}+4\frac{r^4}{A^4}}+2\frac{r^2}{A^2}-\frac{1}{4}\frac{A^2}{R^2}}{C}}, \]
\begin{eqnarray}
\lambda^\prime &=&
\frac{\left(2\frac{r}{R^2}-16\frac{r^3}{A^4}\right)}{\left(1-\frac{r^2}{R^2}+4\frac{r^4}{A^4}\right)},
\end{eqnarray}
\begin{eqnarray}
\nu^\prime&=&\frac{\left(1-\frac{r^2}{R^2}+4\frac{r^4}{A^4}\right)^{-1/2}\left(-\frac{r}{2R^2}+4\frac{r^3}{A^4}\right)+2\frac{r}{A^2}}
{\left(\sqrt{1-\frac{r^2}{R^2}+4\frac{r^4}{A^4}}+2\frac{r^2}{A^2}-\frac{1}{4}\frac{A^2}{R^2}\right)}\nonumber\\
&&2\cot\ln\sqrt{\frac{\sqrt{1-\frac{r^2}{R^2}+4\frac{r^4}{A^4}}+2\frac{r^2}{A^2}-\frac{1}{4}\frac{A^2}{R^2}}{C}}\label{eq4}
\end{eqnarray}
and
\begin{eqnarray}
\nu^{\prime\prime}&=&[\left(\sqrt{1-\frac{r^2}{R^2}+4\frac{r^4}{A^4}}+2\frac{r^2}{A^2}-\frac{1}{4}\frac{A^2}{R^2}\right)^{-1} \times \nonumber
\\&&\{-2\left(1-\frac{r^2}{R^2}+4\frac{r^4}{A^4}\right)^{-3/2}\left(-\frac{r}{2R^2}+4\frac{r^3}{A^4}\right)^{2}\nonumber
\\&&+\left(1-\frac{r^2}{R^2}+4\frac{r^4}{A^4}\right)^{-1/2}\left(-\frac{1}{2R^2}+\frac{12r^2}{A^4}\right)+\frac{2}{A^2}\}\nonumber
\\&&-2\left(\sqrt{1-\frac{r^2}{R^2}+4\frac{r^4}{A^4}}+2\frac{r^2}{A^2}-\frac{1}{4}\frac{A^2}{R^2}\right)^{-2} \times \nonumber
\\&&\left\{\left(1-\frac{r^2}{R^2}+4\frac{r^4}{A^4}\right)^{-1/2}\left(-\frac{r}{2R^2}+4\frac{r^3}{A^4}\right)+\frac{2r}{A^{2}}\right\}^{2}]\nonumber
\\&& \times 2\cot\ln\sqrt{\frac{\sqrt{1-\frac{r^2}{R^2}+4\frac{r^4}{A^4}}+2\frac{r^2}{A^2}-\frac{1}{4}\frac{A^2}{R^2}}{C}}\nonumber
\\&&-2[\frac{\left(1-\frac{r^2}{R^2}+4\frac{r^4}{A^4}\right)^{-1/2}\left(-\frac{r}{2R^2}+4\frac{r^3}{A^4}\right)+\frac{2r}{A^{2}}}
{\sqrt{1-\frac{r^2}{R^2}+4\frac{r^4}{A^4}}+2\frac{r^2}{A^2}-\frac{1}{4}\frac{A^2}{R^2}} \times \nonumber
\\&&cosec\ln\sqrt{\frac{\sqrt{1-\frac{r^2}{R^2}+4\frac{r^4}{A^4}}+2\frac{r^2}{A^2}-\frac{1}{4}\frac{A^2}{R^2}}{C}}]^{2}
\end{eqnarray}

\section{Analysis of Physical behaviour}
In this section we will discuss the following features of the anisotropic strange star :

\subsection{Density and Pressure Behavior of the star}
Now from eqn.(3) and eqn. (6)we get

\begin{eqnarray}
\rho &=& \frac{1}{8\pi}\left(\frac{3}{R^2}-20\frac{r^2}{A^4}\right).\nonumber\\
 Therefore, ~~~~~
\rho_0 &=& \frac{3}{8\pi  R^2}  ,\nonumber\\
\rho_b &=& \frac{1}{8\pi}\left(\frac{3}{R^2}-20\frac{b^2}{A^4}\right),\nonumber
\end{eqnarray}
where we have assumed that $b$ is the radius of the star and $\rho_{0}$ and $\rho_{b}$ is the matter density at center and surface of the star.

Now, we will check, whether at the centre of the star, matter density dominates or not.
Here,we see that
\begin{eqnarray}
\frac{d\rho}{dr} &=& - \frac{5r}{\pi  A^4}< 0,\nonumber\\
\frac{d\rho}{dr} (r=0) &=& 0,\nonumber \\
\frac{d^2 \rho}{dr^2}(r=0) &=& -\frac{5}{\pi  A^4} <0.\nonumber
\end{eqnarray}
Clearly, at the centre of the star, density is maximum and it
decreases
radially outward. \\
Similarly, from Eq.(4), we get
\begin{eqnarray}
\frac{dp_{r}}{dr} &=&  \frac{\frac{8r}{A^4}+\nu^\prime(12\frac{r^2}{A^4}-\frac{1}{r^2}-\frac{1}{R^2})+\nu^{\prime\prime}(4\frac{r^3}{A^4}+\frac{1}{r}-\frac{r}{R^2})}
{8\pi} < 0\nonumber\\
\end{eqnarray}
Now, at the centre(r=0),
\begin{eqnarray}
\frac{dp_{r}}{dr}(r=0) &=& 0\nonumber\\
~~\frac{d^2 p_{r}}{dr^2}(r=0) &=&  < 0\nonumber
\end{eqnarray}
Therefore, at the centre, we also see that the radial pressure is maximum  and it decreases from the centre
towards the boundary. Thus, the energy density and the radial pressure are well behaved
in the interior of the stellar structure. Variations of the
energy-density and two pressures have been shown in Fig.~1 and
Fig.~2, respectively.

The anisotropic parameter $\Delta (r)  = \left(p_t-p_r\right)$ representing the anisotropic
stress is given by Fig.3. The `anisotropy' will be directed outward when $p_{t}>p_{r}$ i.e. $\Delta>0,$ and inward when $p_{t}<p_{r}$ i.e. $\Delta<0.$
It is apparent from the Fig.(3) of our model that a repulsive `anisotropic' force ($\Delta>0$) allows the construction of more massive distributions.
\\
The dimensionless quantity $\omega(r)=\frac{p_{r}+2p_{t}}{3\rho}$
determines a measure of the equation of state. The plot (Fig.4)
for $\omega(r)$ shows that equation of state parameter less than
unity within the  interior of the strange star.

\begin{figure}[htbp]
\centering
\includegraphics[scale=.3]{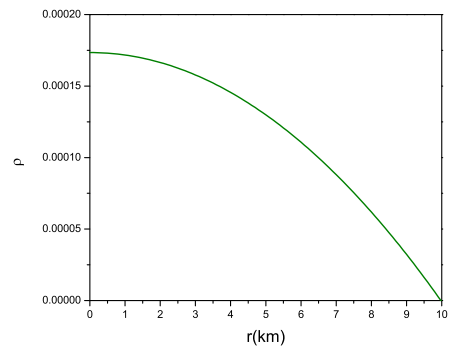}
\caption{Variation of the energy-density($\rho$) at the stellar
interior of the strange star. We have taken the numerical values of
the parameters as $ b=10, R=26.25, A=25.999, C=0.05391.$ }
\label{fig:1}
\end{figure}

\begin{figure}[htbp]
\centering
\includegraphics[scale=.3]{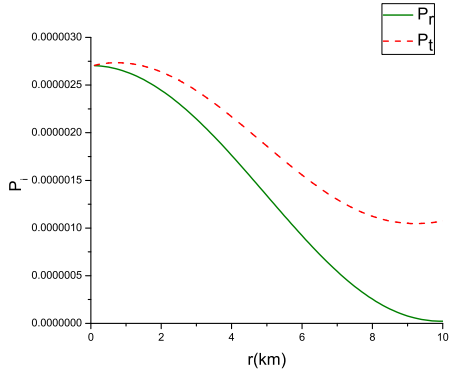}
\caption{Variation of the radial and transverse pressure at the
stellar interior of the strange star. We have taken the numerical
values of the parameters as $ b=10, R=26.25, A=25.999,
C=0.05391.$} \label{fig:2}
\end{figure}

\begin{figure}[htbp]
    \centering
        \includegraphics[scale=.3]{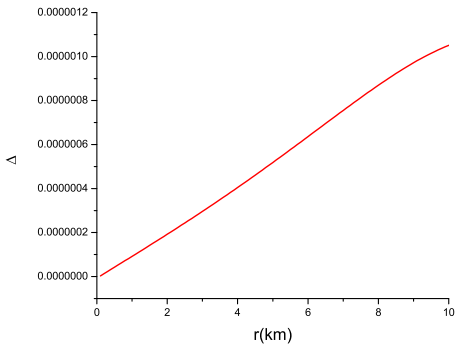}
        \caption{ Effective anisotropic behavior $\Delta (r)$  at the
stellar interior of the strange star. We have taken the numerical
values of the parameters as $ b=10, R=26.25, A=25.999, C=0.05391.$}
    \label{fig:4}
\end{figure}

\begin{figure}[htbp]
\centering
\includegraphics[scale=.3]{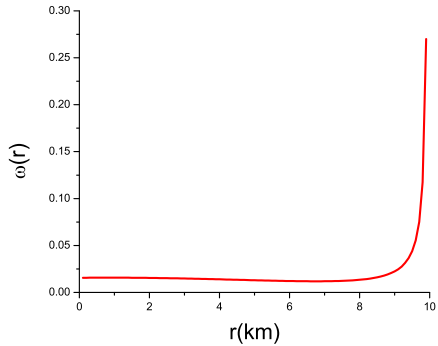}
\caption{Variation of dimensionless quantity
$\omega(r)=\frac{p_{r}+2p_{t}}{3\rho}$ that determines a measure
of the equation of state at the stellar interior of the strange
star. We have taken the numerical values of the parameters as $
b=10, R=26.25, A=25.999, C=0.05391.$} \label{fig:4}
\end{figure}

\subsection{Matching Conditions}
Interior metric of the star should be matched to the Schwarzschild exterior metric at the boundary ($r=b$).
\begin{equation}
ds^2 = - \left(1-\frac{2M}{r}\right)dt^2 +  \left(1-\frac{2M}{r}\right)^{-1}dr^2 +r^2
d\Omega^2  , \label{eq1}
\end{equation}
Assuming the continuity of the metric functions $g_{tt}, ~~g_{rr}$ and $\frac{\partial g_{tt}}{\partial r}$ at the boundary, we get

\begin{eqnarray}
\left(1-\frac{b^2}{R^2}+4\frac{b^4}{A^4}\right) &=& 1 - \frac{2M}{b} \label{eq13}
\end{eqnarray}
and
\begin{equation}
B^2\sin ^{2}\ln \sqrt{\frac{\sqrt{1-\frac{b^2}{R^2}+4\frac{b^4}{A^4}}+2\frac{b^2}{A^2}-\frac{1}{4}\frac{A^2}{R^2}}{C}} =
 \left(1-\frac{2M}{b}\right)  .\label{eq14}
\end{equation}

Now from equation ~(\ref{eq13}) , we get the compactification factor as
\begin{equation}
\frac{M}{b} = \left(\frac{b^2}{2R^2}-2\frac{b^4}{A^4}\right).\label{eq15}
\end{equation}

\subsection{TOV equation}
For an anisotropic fluid distribution, the generalized TOV equation has the form
\begin{equation}
\frac{dp_{r}}{dr} +\frac{1}{2} \nu^\prime\left(\rho+ p_{r}\right)+\frac{2}{r}\left(p_{r}- p_{t}\right)= 0.\label{eq18}
\end{equation}
Following \citep{Leon1993}, we write the above equation as
\begin{equation}
-\frac{M_G\left(\rho+p_{r}\right)}{r^2}e^{\frac{\lambda-\nu}{2}}-\frac{dp_{r}
}{dr}
 +\frac{2}{r}\left(p_{t}-p_{r}\right) = 0, \label{eq19}
\end{equation}
where $M_G(r)$ is the gravitational mass inside a sphere of radius $r$ and is given by
\begin{equation}
M_G(r) = \frac{1}{2}r^2e^{\frac{\nu-\lambda}{2}}\nu^{\prime}.\label{eq20}
\end{equation}
and $e^{\lambda(r)} = (1-\frac{r^2}{R^2}+4\frac{r^4}{A^4} )^{-1}$

which can easily be derived from the Tolman-Whittaker formula and the Einstein's field equations. The modified  TOV equation describes the equilibrium condition for the strange star subject to
effective gravitational($F_g$) and effective hydrostatic($F_h$) plus another force due to the effective anisotropic($F_a$)
 nature of the stellar object as
\begin{equation}
F_g+ F_h + F_a = 0,\label{eq21}
\end{equation}
where the force components are given by
\begin{eqnarray}
F_g &=& -\frac{1}{2}\nu^{\prime}\left(\rho+p_{r}\right),\label{eq22}\\
F_h &=& -\frac{dp_{r}}{dr} \label{eq23}\\  ~~ F_a &=&
\frac{2}{r}\left(p_{t} -p_{r}\right).\label{eq24}
\end{eqnarray}
We plot ( Fig. 5 )   the behaviors of pressure anisotropy,
gravitational and hydrostatic forces in  the   interior region
which shows sharply   that the static equilibrium configurations
do exist due to the combined effect of pressure anisotropy,
gravitational and hydrostatic forces.

\begin{figure}[htbp]
    \centering
        \includegraphics[scale=.3]{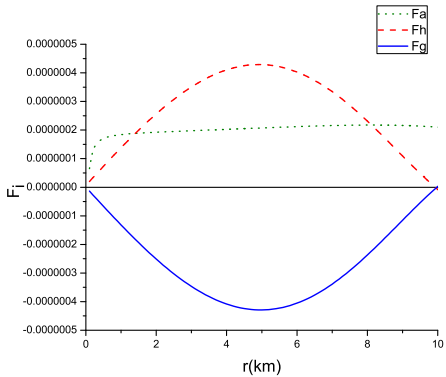}
        \caption{ Behaviors of pressure anisotropy, gravitational and hydrostatic forces at the
stellar interior of strange star. We have taken the numerical values
of the parameters as $ b=10, R=26.25, A=25.999, C=0.05391.$}
    \label{fig:5}
\end{figure}

\subsection{Energy conditions}
All the energy conditions, namely, null energy condition(NEC), weak energy condition(WEC), strong energy condition(SEC)
and dominant energy condition(DEC), are satisfied at the centre ($r=0$).\\
(i) NEC: $p_{0}+\rho_{0}\geq0$ ,\\
(ii) WEC: $p_{0}+\rho_{0}\geq0$  , $~~\rho_{0}\geq0$  ,\\
(iii) SEC: $p_{0}+\rho_{0}\geq0$  ,$~~~~3p_{0}+\rho_{0}\geq0$ ,\\
(iv) DEC: $\rho_{0} > |p_{0}| $.\\
We have assumed  the numerical values of the parameters $ R=26.25,
A=25.999, C=0.05391 $ to calculate above    energy conditions.
\subsection{Stability}
For a physically acceptable model, one expects that the velocity of
sound should be within the range  $0 \leq  v_s^2=(\frac{dp}{d\rho})
\leq 1$\citep{Herrera1992,Abreu2007}. According to Herrera's \citep{Herrera1992} cracking
(or overturning) condition : The region  for which radial speed of
sound is greater than the transverse speed of sound is a
potentially stable
region.\\
  In our case(anisotropic strange stars),
we plot the radial and transverse sound speeds in Fig.6 and
observe that these parameters satisfy the inequalities $0\leq
v_{sr}^2 \leq 1$ and $0\leq v_{st}^2 \leq 1$ everywhere within
the stellar object. We also note that $v^2_{st}-v^2_{sr}\leq 1$.
Since, $0\leq v_{sr}^2 \leq 1$ and $0\leq v_{st}^2 \leq 1$,
therefore, $\mid v_{st}^2 - v_{sr}^2 \mid \leq 1 $. In Fig.7, we
have plotted  $\mid v_{st}^2 - v_{sr}^2 \mid$. We  notice  that
$v^2_{st}<v^2_{sr} $ throughout the interior region. In other
words,   $v^2_{st}<v^2_{sr} $  keeps the same sign everywhere
within the matter distribution i.e. no cracking will occur. These
results show that our anisotropic compact stars model is stable.

\begin{figure}[htbp]
   \centering
        \includegraphics[scale=.3]{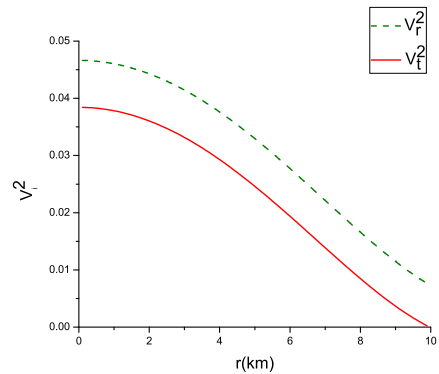}
       \caption{ Variation of the radial and transverse sound speed of the strange star. We have taken the
numerical values of the parameters as $ b=10, R=26.25, A=25.999,
C=0.05391.$}
    \label{fig:6}
\end{figure}

\begin{figure}[htbp]
    \centering
        \includegraphics[scale=.3]{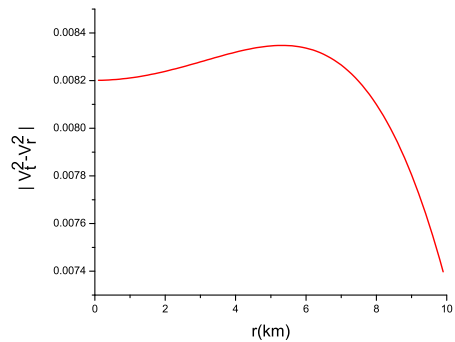}
       \caption{ Variation of  $|v_{st}^2 - v_{sr}^2|$ of the strange star.}
    \label{fig:7}
\end{figure}
\subsection{Mass-Radius relation and Surface redshift}
In this section, we study the maximum allowable mass-radius ratio
in our model. According to Buchdahl \citep{Buchdahl1959}, for a
static spherically symmetric perfect fluid allowable mass-radius
ratio is given by $\frac{2 Mass}{Radius} < \frac{8}{9}$.
Mak\citep{Mak2001} also gave more generalized expression. In our
model the gravitational mass in terms of the energy density
$\rho$ can be expressed as
\begin{equation}
\label{eq34}
 M=4\pi\int^{b}_{0} \rho~~ r^2 dr =
 \frac{b}{2}\left[\frac{b^2}{R^2}-4\frac{b^4}{A^4}\right]
\end{equation}

 The compactness of the star is given by
\begin{equation}
\label{eq35} u= \frac{ M(b)} {b}=
 \frac{1}{2}\left[\frac{b^2}{R^2}-4\frac{b^4}{A^4}\right]
\end{equation}
The nature of the Mass and Compactness of the star from the
centre are shown   in Fig. 8 and Fig.9.
\begin{figure}[htbp]
\centering
\includegraphics[scale=.3]{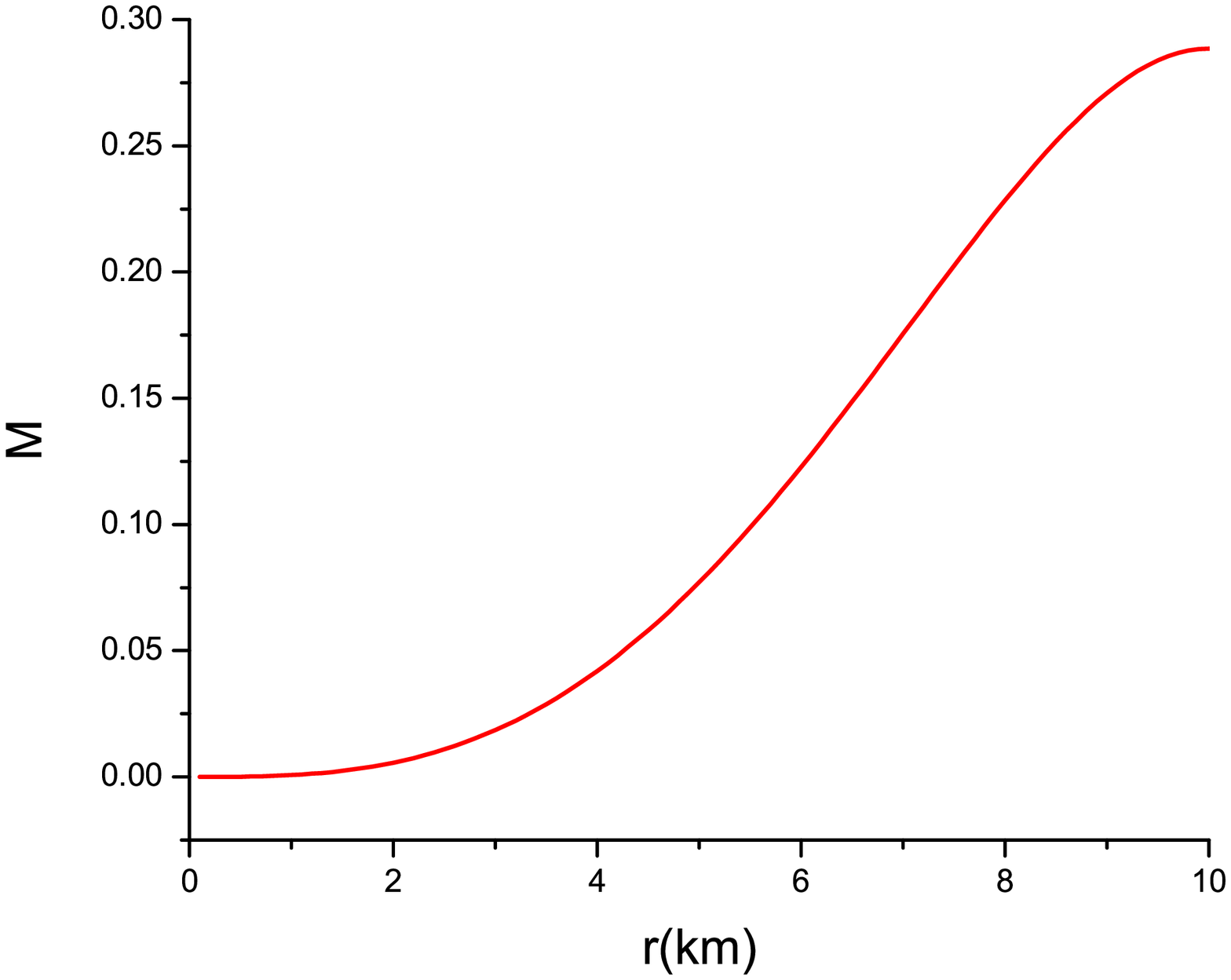}
\caption{Variation of the Mass function of the strange star. We
have taken the numerical values of the parameters as $ b=10,
R=26.25, A=25.999, C=0.05391.$} \label{fig:8}
\end{figure}

\begin{figure}[htbp]
\centering
\includegraphics[scale=.3]{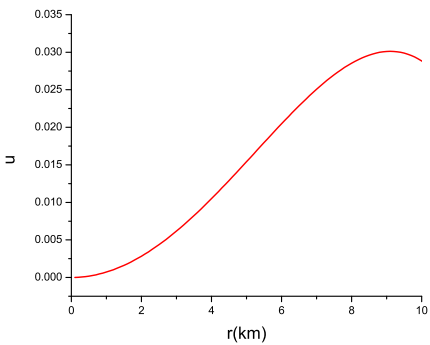}
\caption{Variation of Compactness of the strange star. We have
taken the numerical values of the parameters as $ b=10, R=26.25,
A=25.999, C=0.05391.$ } \label{fig:9}
\end{figure}

The surface redshift ($Z_s$) corresponding to the above
compactness ($u$) is obtained as
\begin{equation}
\label{eq36} 1+Z_s= \left[ 1-(2 u )\right]^{-\frac{1}{2}} ,
\end{equation}
where
\begin{equation}
\label{eq37} Z_s= \frac{1}{\sqrt{1-\frac{b^2}{R^2}+4\frac{b^4}{A^4}}}-1
\end{equation}
Thus, the maximum surface redshift for the anisotropic strange
stars of different radius could be found very easily from the
Fig. 10. We calculate the maximum surface redshift for our
configuration using the numerical values of the parameters  as $
b=9.5, R=16.9, A=24.18$ and we get $Z_s=0.375$. The nature of
surface redshift  of the star is shown   in Fig. 10.

\begin{figure}[htbp]
\centering
\includegraphics[scale=.3]{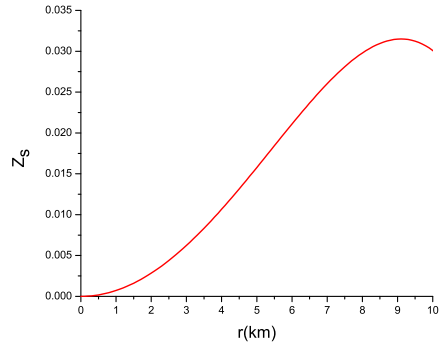}
\caption{Variation of Red-shift function of the strange star. We
have taken the numerical values of the parameters as $ b=10,
R=26.25, A=25.999, C=0.05391.$} \label{fig:10}
\end{figure}

\section{Conclusion}
In this work we have investigated the nature of anisotropic strange
stars in the low-mass X-ray binary 4U 1820-30 by taking following considerations : (a) The stars are
anisotropic in nature i.e. $p_{r}\neq p_{t}$. (b) The space-time
of the strange stars can be described by Tolman VII metric.   \\

The results are quite interesting, which are as follows: (i) Though
the radial pressure$(p_{r})$ vanishes at the boundary $(r=b)$,
tangential pressure$(p_{t})$ does not. However, at the centre of the
star, it's anisotropic behavior vanishes. (ii) Our model is well
stable according to Herrera
 stability condition \citep{Herrera1992}. (iii) From mass-radius relation, any interior features of the star can be evaluated.\\

Therefore, our overall observations of anisotropic strange stars
under Tolman VII metric satisfies all physical requirements of a
stable star.
\\
 It is to be noted  that while solving Einstein's
equations as well as for plotting, we have set c=G=1.
  Now, plugging G and c into relevant equations, the
  values of the central density and  surface density of our strange star turn out to be $ \rho_0 =
 0 .55 ~\times ~ 10^{15}~ gm ~cm^{-3}$ and $ \rho_b = 0.27 ~\times ~ 10^{15}~ gm ~cm^{-3}$ for
the numerical values of the parameters as $ b=9.5, R=16.9, A=24.18$.
Also, the mass of our strange star is calculated as $ 1.01M_\odot $.
Interestingly, we observe that  the measurement of the mass, radius
and central density  of our strange star are almost consistent with
the strange star in the low-mass X-ray binary 4U 1820-30
\citep{Guver2010}.

 Recently, Cackett et al. \cite{Cackett} reported that the
gravitational redshift of strange star in the low-mass X-ray
binary 4U 1820-30, based on the modeling of the relativistically
broadened iron line in the X-ray spectrum of the source observed
with Suzaku  is $Z_s= 0.43$. The surface redshift of our strange
star with  radius $9.5$ km turns out to be $0.375$. This indicates
that the measurement of redshift  of our strange star  is nearly
reliable with the strange star in the low-mass X-ray binary 4U
1820-30.

Finally, we conclude by pointing that spacetime comprising Tolman
VII metric with anisotropy may be used to construct a suitable model
of a strange star in the low-mass X-ray binary 4U 1820-30.

\section*{Acknowledgments} MK, FR and SMH gratefully acknowledge support
 from IUCAA, Pune, India under Visiting Associateship under which a part
  of this work was carried out.  FR is also thankful to UGC, for providing
financial support under research award scheme.  We are grateful to
the referee for his valuable suggestions.

\end{document}